\documentclass[aps,graphicx,6pt,showkeys,showpacs]{revtex4}
\usepackage{amsmath}
\usepackage{amscd}
\usepackage{graphicx}
\usepackage{subfigure}
\usepackage{appendix}

\begin{document}

\title[Short Title]{Fast preparation of W states with superconducting quantum interference devices by using dressed states}

\author{Yi-Hao Kang$^{1}$}
\author{Ye-Hong Chen$^{1}$}
\author{Zhi-Cheng Shi$^{1}$}
\author{Jie Song$^{2}$}
\author{Yan Xia$^{1,}$\footnote{E-mail: xia-208@163.com}}

\affiliation{$^{1}$Department of Physics, Fuzhou University, Fuzhou 350002, China\\
             $^{2}$Department of Physics, Harbin Institute of Technology, Harbin 150001, China}

\begin{abstract}
In this paper, we propose a protocol to prepare W states with
superconducting quantum interference devices (SQUID) by using
dressed states. Through choosing a set of dressed states suitably,
the protocol can be used to accelerate the adiabatic passages while
additional couplings are unnecessary. Moreover, we can optimize the
evolution of the system with the restraint to the populations of the
intermediate states by choosing suitable controlled parameters.
Numerical simulations show that the protocol is robust against the
parameter variations and decoherence mechanisms. Furthermore, the
protocol is faster and more robust against the dephasing, compared
with that by the adiabatic passages. As for the Rabi frequencies of
pulses designed by the method, they can be expressed by the linear
superpositions of Gaussian functions, which does not increase
difficulties to the experiments. In addition, the protocol could be
controlled and manipulated easily in experiments with a circuit
quantum electrodynamics system.


\end{abstract}

\maketitle

\section{Introduction}

Executing computation and communication tasks
\cite{LeePRL96,YangPRA82,AmniatPRA71,SaffmanRMP82} in quantum
information processing (QIP) are very attractive in recent years,
since these tasks can be accurately completed with suitable boundary
condition of time-dependent interactions. For example, based on the
idea of guiding the evolution of the system ``riding'' the adiabatic
eigenstates from its initial state to the target state, adiabatic
methods have been proposed, and widely used successfully in many
research fields, such as laser cooling and atom optics
\cite{RuschhauptPRA73}, metrology \cite{PereiraPRL89},
interferometry \cite{WeitzPRL73}, chemical reaction dynamics
\cite{KralRMP79}, cavity quantum electrodynamics
\cite{BergmannRMP70}, etc.. The most famous examples of adiabatic
methods are the stimulated Raman adiabatic passages (STIRAP) and its
variants \cite{FewellAJP50,BergmannRMP70,VitanovARPC52,KralRMP79},
which have shown many advantages. For instance, the protocols with
the STIRAP have great robustness against pulse area and timing
errors. Moreover, when the system stays in the instantaneous ground
state of its time-dependent Hamiltonian during the whole evolution
process under an adiabatic control, the populations of the lossy
intermediate states can be restrained so that the dissipation caused
by decoherence, noise and losses can be repressed. Although the
adiabatic passages hold several advantages, the methods with STIRAP
require the system being restricted by the adiabatic condition,
which may greatly reduce the evolution speed of the system and make
the system suffering more from the dissipation of its initial state
and target state. For example, as shown in Refs.
\cite{WeiQIP14,WuQIP}, by using STIRAP to create entanglement, the
fidelities of obtaining the target states are very sensitive to the
dephasing due to a long time evolution. It is generally known that
in the field of quantum computing and quantum-information
processing, the speed and precision are two primary factors.
Therefore, in order to drive a system from a given initial state to
a prescribed final state in a shorter time without losing the
robustness property, a new sort of technique called ``Shortcuts to
adiabatic passages'' (STAP)
\cite{DemirplakJPCA107,DemirplakJCP129,TorronteguiAAMOP62,BerryJPA42,ChenPRL105,CampoPRL111,ChenPRA83,MugaJPB42,ChenPRL104,CampoSR2}
has been put forward.

The STAP aims at leading an adiabatic-like way between the system's
initial state and the target state while the adiabatic condition is
completely broken so that the evolution of the system can be
accelerated a lot. Moreover, when suitable boundary condition of
time-dependent interactions are set, the robustness of STAP against
parameter variations and decoherence mechanisms is also quite nice.
Because of the attractive advantages, the STAP has been applied in
many kinds of research fields, e.g., ``fast cold-atom'', ``fast ion
transport'', ``fast quantum information processing'', ``fast
wave-packet splitting'', ``fast expansion'', and so on
\cite{TorronteguiPRA83,MugaJPB43,TorronteguiPRA85,MasudaPRA84,yehongPRA91,ChenPRA82,yehongOC140,SchaffNJP13,ChenPRA84,TorronteguiNJP14,CampoPRA84,CampoPRL109,Anarxiv,DuNC7,ZhangPRL110,CampoEPL96,RuschhauptNJP14,SchaffPRA82,SchaffEPL93,ChenPRA86,chenzhenSR6,lumeiPRA89,yehongPRA89,SongNJP6,yehongSR5,xiaobinQIP14,wujiangQIP15,SantosSR5,SantosPRA93,HenPRA91,SarandyQIP3,Coulamyarxiv,RamsarXiv,DeffnerPRX4,Campomeeting}.
Among these works
\cite{DemirplakJPCA107,DemirplakJCP129,TorronteguiAAMOP62,yehongOC140,BerryJPA42,ChenPRL105,CampoPRL111,ChenPRA83,MugaJPB42,ChenPRL104,CampoPRA84,CampoPRL109,Anarxiv,DuNC7,ZhangPRL110,CampoSR2,TorronteguiPRA83,MugaJPB43,TorronteguiPRA85,MasudaPRA84,yehongPRA91,ChenPRA82,SchaffNJP13,ChenPRA84,TorronteguiNJP14,CampoEPL96,RuschhauptNJP14,SchaffPRA82,SchaffEPL93,ChenPRA86,chenzhenSR6,lumeiPRA89,yehongPRA89,SongNJP6,yehongSR5,xiaobinQIP14,wujiangQIP15,SantosSR5,SantosPRA93,HenPRA91,SarandyQIP3,Coulamyarxiv,RamsarXiv,DeffnerPRX4,Campomeeting},
shortcut protocols
\cite{BerryJPA42,ChenPRL105,CampoPRL111,ChenPRA83,chenzhenSR6,SongNJP6,yehongSR5,xiaobinQIP14,wujiangQIP15,yehongOC140}
with the method named ``transitionless quantum driving'' (TQD) are
interesting. In these protocols
\cite{BerryJPA42,ChenPRL105,CampoPRL111,ChenPRA83,chenzhenSR6,SongNJP6,yehongSR5,xiaobinQIP14,wujiangQIP15,yehongOC140},
modifications of original Hamiltonians could be constructed to
compensate for nonadiabatic errors by adding ``counter-diabatic
driving'' (CDD) terms with TQD. However, as indicated in Ref.
\cite{BaksicPRL116}, the CDD terms sometimes paly roles as either
direct couplings between the initial state and the target state
\cite{ChenPRL105,GiannelliPRA89,MasudaJPCA119} or couplings not
available in the original Hamiltonian \cite{BasonNP8}. It has been
shown in some previous protocols
\cite{yehongOC140,chenzhenSR6,yehongSR5,xiaobinQIP14,wujiangQIP15}
that, a direct coupling between the initial state and the target
state may be hard to be realized in several cases, such as the
special one-photon 1-3 pulse (the microwave field) for an atom
transition. Therefore, many other interesting approaches
\cite{GaraotPRA89,OpatrnyNJP16,SaberiPRA90,TorronteguiPRA89,TorosovPRA87,TorosovPRA89,yehongPRA93,KangSR6,IbanezPRA87,IbanezPRL109,SongPRA93,BaksicPRL116}
have been presented to construct STAP and avoid the issues caused by
TQD. For example, Torrontegui \emph{et al.} \cite{TorronteguiPRA89}
have used the dynamical symmetry of the Hamiltonian to find
alternative Hamiltonians that achieved the same goals as speed-up
protocols via Lie transforms without directly using the
counterdiabatic Hamiltonian. Ib\'{a}\~{n}ez \emph{et al.}
\cite{IbanezPRL109} have suggested to use iterative interaction
pictures (also called the ``multiple Schr\"{o}dinger pictures'') to
obtain Hamiltonians with physically feasible structure for quantum
systems. They have also studied the capabilities and limitations of
superadiabatic iterations to construct a sequence of shortcuts to
adiabaticity by iterative interaction pictures \cite{IbanezPRA87}.
Subsequently, the method with multiple Schr\"{o}dinger pictures has
been expanded by Song \emph{et al.} \cite{SongPRA93} to a
three-level system. They have shown an interesting result that the
Hamiltonian in the second iteration of the interaction pictures has
the same form as the Hamiltonian in the original Schr\"{o}dinger
picture \cite{SongPRA93}. Recently, Baksic \emph{et al.}
\cite{BaksicPRL116} have proposed an interesting protocol about
significantly speeding up adiabatic state transfers by using dressed
states. Moreover, they have indicated in their article
\cite{BaksicPRL116} that the populations of the intermediate states
can be controlled by choosing one of the controlled parameters and
such control is unable in the protocols with superadiabatic
iterations. This result is quite attractive, since one can decrease
the populations of the intermediate states by adjusting the
corresponding parameters in order to reduce the dissipation of the
intermediate states and improve the fidelity of obtaining the target
state. Considering the advantages of the method by using dressed
states, it is worthwhile to dig out the applications of this method
for QIP in various physics systems.

On the other hand, it has been reported in the recent developments
in circuit quantum electrodynamics, superconducting devices
(including single Cooper pair boxes, Josephson junctions, and
superconducting quantum interference devices (SQUIDs)) have a
natural superiority for their scalability to be regarded as very
prospective candidates to implement QIP
\cite{MakhlinRMP73,VionSci296,YangPRA67,YangPRL92,YangPRA74,YangPRA86,YangPRA87,YangSR4,NakamuraNat398,SteinbachPRL87,MartinisPRL63,RousePRL75,WalSci290,HanPRL76,FriedmanNat406}.
Superconducting qubits are relatively easy to scale up and have a
long decoherence time \cite{VionSci296,YuSci296,ChiorescuNature431}.
Moreover, using SQUID qubits in cavity quantum electrodynamics (QED)
have several advantages. For example, the positions of SQUID qubits
in a cavity are fixed while for cavity-atom systems it remains a
significant technical challenge to control the center of mass motion
of a neutral atom \cite{YangPRL92,YangPRA67}. Besides, by changing
local bias fields or designing suitable variations, level structure
of every individual SQUID qubit can be adjusted readily
\cite{YangPRA67}. Furthermore, when SQUID qubits are embedded in a
cavity, the strong-coupling limit of the cavity QED can be easily
realized while for atoms in a cavity, that is difficult to be
achieved \cite{YangPRL92}. Therefore, SQUID qubits are attractive
tools for implement quantum information tasks.

Combining the advantages of the method with dressed states
\cite{BaksicPRL116} and SQUID qubits, we investigate the
entanglement preparation in the present protocol. Considering the
importance of W states in both examining quantum nonlocality
\cite{DurPRA62} and implementing quantum information tasks
\cite{JungPRA78,KarlssonPRA58}, we prepare W states for three SQUID
qubits by using dressed states as an example. In this protocol,
laser pulses can be designed so that a W state of three SQUID qubits
can be obtained with high speed without using any additional
couplings. Besides, the Rabi frequencies of pulses designed by the
method with dressed states could be realized without challenges in
experiments since they can be expressed by the linear superpositions
of Gaussian functions. By selecting suitable controlled parameters,
the populations of the intermediate states can be restrained, hence
the system will suffer less from dissipation of intermediate states.
Numerical simulations demonstrate that the protocol is robust
against the parameter variations and decoherence mechanisms.
Different from the protocol for generating W states with the
adiabatic passages in Ref. \cite{DengPRA74}, in this paper, through
choosing a set of dressed states suitably, the protocol can be used
to accelerate the adiabatic passages while additional couplings are
unnecessary. So, the W state can be generated faster than that in
Ref. \cite{DengPRA74}. On the other hand, limited by the adiabatic
condition, the W state generation in Ref. \cite{DengPRA74} is more
sensitive to the dephasing. On the contrary, since the W state can
be fast generated here, the present protocol is much more robust
against the dephasing. Therefore, the present protocol is more
feasible for experimental realization.

The article is organized as follows. In Sec. II, we will review the
method to accelerate the adiabatic passages by using dressed states
proposed in Ref. \cite{BaksicPRL116}. In Sec. III, we will describe
how to prepare W state of three SQUID qubits by using dressed
states. In Sec. IV, we will investigate the performance of the
protocol via numerical simulations. And finally, the conclusion will
be given in Sec. V.

\section{Accelerating the adiabatic passages by using dressed states}

In this section, we would like to review the method to accelerate
the adiabatic passages by using dressed states proposed in Ref.
\cite{BaksicPRL116}. Firstly, we define a picture transformation
$U(t)=\sum\limits_n|\varphi_n(t)\rangle\langle n|$, where,
$\{|\varphi_n(t)\rangle\}$ are the instantaneous eigenstates of the
original Hamiltonian $H_0(t)$ corresponding to the eigenvalues
$\{E_n(t)\}$, and $\{|n\rangle\}$ are a set of time-independent
states. In adiabatic picture, the Hamiltonian becomes
\begin{eqnarray}\label{e1}
H_{ad}(t)=U^{\dag}(t)H_0(t)U(t)+W(t)=\sum\limits_nE_n(t)|n\rangle\langle
n|-iU^{\dag}(t)\dot{U}(t),
\end{eqnarray}
in which $W(t)=-iU^{\dag}(t)\dot{U}(t)$ generically has off-diagonal
matrix elements connecting the various instantaneous eigenstates of
$H_0(t)$ and causing nonadiabatic errors. In order to correct the
nonadiabatic errors, a correction Hamiltonian $H_{co}(t)$ is
introduced such that the modified Hamiltonian
$H'(t)=H_0(t)+H_{co}(t)$. Therefore, in adiabatic picture, the
modified Hamiltonian becomes
\begin{eqnarray}\label{e2}
H'_{ad}(t)&&=U^{\dag}(t)H_0(t)U(t)+U^{\dag}(t)H_{co}(t)U(t)+W(t)\cr\cr&&
=\sum\limits_nE_n(t)|n\rangle\langle
n|+U^{\dag}(t)H_{co}(t)U(t)-iU^{\dag}(t)\dot{U}(t)\cr\cr&&
=H_{ad}(t)+U^{\dag}(t)H_{co}(t)U(t).
\end{eqnarray}

Secondly, we define another picture transformation
$V(t)=\sum\limits_n|\tilde{\varphi}_n(t)\rangle\langle n|$, where
$\{|\tilde{\varphi}_n(t)\rangle\}$ are a set of dressed states.
Assuming that the initial time is $t_i$ and the final time is $t_f$,
the unitary operator $V(t)$ should satisfy the condition
$V(t_i)=V(t_f)=1$. Then, we move from the adiabatic picture to the
new picture called ``dressed-state picture''. $H'_{ad}(t)$ in
adiabatic picture will become
\begin{eqnarray}\label{e3}
H'_{V}(t)=V^{\dag}(t)H_{ad}(t)V(t)+V^{\dag}(t)U^{\dag}(t)H_{co}(t)U(t)V(t)-iV^{\dag}(t)\dot{V}(t).
\end{eqnarray}
Afterwards, $H_{co}(t)$ should be carefully designed so that the
modified Hamiltonian $H'_{V}$ and the dressed states
$\{|\tilde{\varphi}_n(t)\rangle\}$ satisfy
$\langle\tilde{\varphi}_m(t)|H'_{V}(t)|\tilde{\varphi}_n(t)\rangle=0$
$(m\neq n)$, i.e., $H_{co}(t)$ has to be designed for canceling the
unwanted off-diagonal elements in $H_V(t)$.

\section{Fast preparation of W states for three SQUID qubits by using dressed states}

Let us investigate the entanglement preparation with SQUID qubits by
using dressed states. As an example, we will expound how to prepare
W states of three SQUID qubits by using dressed states. The SQUID
qubits considered here are rf SQUID qubits. Each SQUID qubit
consisting of a Josephson tunnel junction in a superconducting loop.
The Hamiltonian of each rf SQUID qubit can be described as
\cite{YangPRA67,YangPRL92}
\begin{eqnarray}\label{e4}
H_{s}(t)=\frac{Q^2}{2C}+\frac{(\Phi-\Phi_x)^2}{2L}-E_J\cos(2\pi\frac{\Phi}{\Phi_0}),
\end{eqnarray}
in which $C$ is the junction capacitance and $L$ is the loop
inductance, $Q$ is the total charge on the capacitor, $\Phi$ is the
magnetic flux threading the loop, $\Phi_x$ is the external flux
applied to the ring, $\Phi_0=h/2e$ is the flux quantum,
$E_J=I_c\Phi_0/2\pi$ is the Josephson energy with $I_c$ being the
critical current of the junction. We consider that there are four
SQUID qubits, $SQUID_1$, $SQUID_2$, $SQUID_3$ and $SQUID_4$, coupled
to a single-mode microwave cavity field.
\begin{figure}
\scalebox{0.6}{\includegraphics[scale=1]{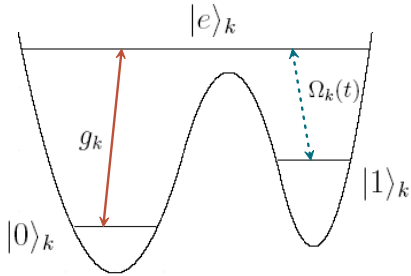}} \caption{The
level configuration for $SQUID_k$ ($k=1,2,3,4$).}
\end{figure}
As shown in Fig. 1, $SQUID_k$ ($k=1,2,3,4$) has the $\Lambda$-type
configuration formed, that is, an excited level $|e\rangle_k$ and
two lowest levels $|0\rangle_k$ and $|1\rangle_k$. The classical
field with Rabi frequency $\Omega_k(t)$ drives the transition
resonantly between the levels $|e\rangle_k$ and $|1\rangle_k$, while
the cavity field couples resonantly to the levels $|0\rangle_k$ and
$|e\rangle_k$ with coupling constant $g_k$. $\Omega_k(t)$ and $g_k$
are given in Refs. \cite{YangPRA67,YangPRL92} as
\begin{eqnarray}\label{e5}
&&g_k=\frac{1}{L_k}\sqrt{\frac{\omega_c}{2\mu_0\hbar}}\langle0|\Phi|e\rangle_k\int_{S_k}\mathbf{B}_c^k(\mathbf{r})\cdot
d\mathbf{S},\cr\cr&&
\Omega_k(t)=\frac{1}{2L_k\hbar}\langle1|\Phi|e\rangle_k\int_{S_k}\mathbf{B}_{\mu
w}^k(\mathbf{r},t)\cdot d\mathbf{S},
\end{eqnarray}
where, $S_k$ is surface bounded by the loop of the $SQUID_k$, $L_k$
is the loop inductance of $SQUID_k$, $\omega_c$ is the cavity
frequency, $\mathbf{B}_c^k(\mathbf{r})$ and $\mathbf{B}_{\mu
w}^k(\mathbf{r},t)$ are the magnetic components of the cavity mode
and the classical microwave in the superconducting loop of the
$SQUID_k$. The Hamiltonian of the system in the interaction picture
with the rotating-wave approximation can be described as ($\hbar=1$)
\begin{eqnarray}\label{e8}
&&H_{I}(t)=H_c+H_m(t),\cr\cr&&
H_c=\sum\limits_{k=1}^{4}g_k|e\rangle_k\langle0|a+H.c.,\cr\cr&&
H_m(t)=\sum\limits_{k=1}^{4}\Omega_k(t)|e\rangle_k\langle1|+H.c.,
\end{eqnarray}
in which $a$ denotes photon annihilation operator of the cavity
mode. For simplicity, we set $g_1=g_2=g_3=g$ and $g_4=\sqrt{3}g$,
which can be realized by adjusting location or parameters of
$SQUID_k$ (e.g. $L_k$ and $S_k$). Moreover, we assume the system is
initially in state
$|\Psi(0)\rangle=|0\rangle_1|0\rangle_2|0\rangle_3|1\rangle_4|0\rangle_c$
($|0\rangle_c$ and $|1\rangle_c$ are the vacuum state and one-photon
state of the cavity mode, respectively). Defining the excited number
operator of the system as $N_e=\sum\limits_k(|e\rangle_k\langle
e|+|1\rangle_k\langle 1|)+a^{\dag}a$, one can obtain that
$[N_e,H_I]=0$ and $\langle\Psi(0)|N_e|\Psi(0)\rangle=1$. Therefore,
the evolution of the system will stay in the one-excited sub-system
spanned by
\begin{eqnarray}\label{e9}
&&|\psi_1\rangle=|0\rangle_1|0\rangle_2|0\rangle_3|1\rangle_4|0\rangle_c,\
|\psi_2\rangle=|0\rangle_1|0\rangle_2|0\rangle_3|e\rangle_4|0\rangle_c,\
|\psi_3\rangle=|0\rangle_1|0\rangle_2|0\rangle_3|0\rangle_4|1\rangle_c,\cr\cr&&
|\psi_4\rangle=|e\rangle_1|0\rangle_2|0\rangle_3|0\rangle_4|0\rangle_c,\
|\psi_5\rangle=|0\rangle_1|e\rangle_2|0\rangle_3|0\rangle_4|0\rangle_c,\
|\psi_6\rangle=|0\rangle_1|0\rangle_2|e\rangle_3|0\rangle_4|0\rangle_c,\cr\cr&&
|\psi_7\rangle=|1\rangle_1|0\rangle_2|0\rangle_3|0\rangle_4|0\rangle_c,\
|\psi_8\rangle=|0\rangle_1|1\rangle_2|0\rangle_3|0\rangle_4|0\rangle_c,\
|\psi_9\rangle=|0\rangle_1|0\rangle_2|1\rangle_3|0\rangle_4|0\rangle_c.
\end{eqnarray}
Here, we would like to prepare the W state
$|W\rangle=\frac{1}{\sqrt{3}}(|\phi_7\rangle+|\phi_8\rangle+|\phi_9\rangle)$
of $SQUID_1$, $SQUID_2$ and $SQUID_3$. $SQUID_4$ is used to provide
a photon to the cavity. Then, we rewrite $H_c$ in this one-excited
subspace as
$H_c=\sqrt{3}g|\psi_2\rangle\langle\psi_3|+g(|\psi_4\rangle+|\psi_5\rangle+|\psi_6\rangle)\langle\psi_3|+H.c.$.
Assuming
$|\varsigma\rangle=\frac{1}{\sqrt{3}}(|\psi_4\rangle+|\psi_5\rangle+|\psi_6\rangle)$,
we have
$H_c=\sqrt{3}g|\psi_1\rangle\langle\psi_2|+\sqrt{3}g|\varsigma\rangle\langle\psi_2|+H.c.$.
The eigenstates of $H_c$ are calculated in the following
\begin{eqnarray}\label{e10}
&&|\phi_0\rangle=\frac{1}{\sqrt{2}}(-|\psi_2\rangle+|\varsigma\rangle),\cr\cr&&
|\phi_1\rangle=\frac{1}{2}(|\psi_2\rangle+\sqrt{2}|\psi_3\rangle+|\varsigma\rangle),\cr\cr&&
|\phi_2\rangle=\frac{1}{2}(|\psi_2\rangle-\sqrt{2}|\psi_3\rangle+|\varsigma\rangle),
\end{eqnarray}
with eigenvalues $E_0=0$, $E_1=\sqrt{6}g$, $E_3=-\sqrt{6}g$,
respectively. For simplicity, we assume that
$\Omega_1(t)=\Omega_2(t)=\Omega_3(t)=\sqrt{2}\Omega_a(t)$ and
$\Omega_4(t)=\sqrt{2}\Omega_b(t)$. By adding the condition
$\Omega_a,\Omega_b\ll g$, the effective Hamiltonian of the system
can be given by
\begin{eqnarray}\label{e11}
H_{eff}(t)&&=\frac{\Omega_a(t)}{\sqrt{3}}(|\psi_7\rangle+|\psi_8\rangle+|\psi_9\rangle)\langle\phi_0|-\Omega_b(t)|\psi_1\rangle\langle\phi_0|+H.c.,\cr\cr&&
=\Omega_a(t)|W\rangle\langle\phi_0|-\Omega_b(t)|\psi_1\rangle\langle\phi_0|+H.c..
\end{eqnarray}

Assuming $\Omega_a(t)=\Omega(t)\cos\theta(t)$ and
$\Omega_b(t)=\Omega(t)\sin\theta(t)$, the three instantaneous
eigenstates of $H_{eff}(t)$ can be described as
\begin{eqnarray}\label{e12}
&&|\varphi_0(t)\rangle=\cos\theta|\psi_1\rangle+\sin\theta|W\rangle,\cr\cr&&
|\varphi_+(t)\rangle=\frac{1}{\sqrt{2}}(\sin\theta|\psi_1\rangle+|\phi_0\rangle-\cos\theta|W\rangle),\cr\cr&&
|\varphi_-(t)\rangle=\frac{1}{\sqrt{2}}(\sin\theta|\psi_1\rangle-|\phi_0\rangle-\cos\theta|W\rangle),
\end{eqnarray}
with eigenvalues $\epsilon_0=0$, $\epsilon_+(t)=\Omega(t)$,
$\epsilon_-(t)=-\Omega(t)$, respectively. A general adiabatic state
transfer from the initial state $|\psi_1\rangle$ to the target state
$|W\rangle$ can be performed via $|\varphi_0(t)\rangle$ with
boundary condition $\theta(0)=0$ and $\theta(T)=\pi/2$. To speed up
the evolution using dressed states, we firstly go into the adiabatic
picture. By using picture transformation
$U(t)=\sum\limits_{n=0,+,-}|\varphi_n(t)\rangle\langle n|$, the
Hamiltonian in adiabatic picture is
\begin{eqnarray}\label{e13}
H_{ad}(t)=\Omega(t)M_z+\dot{\theta}(t)M_y,
\end{eqnarray}
where
\begin{eqnarray}\label{e14}
M_x=\frac{1}{\sqrt{2}}\left[%
\begin{array}{ccc}
  0 &\  -1 &\  1 \\
  -1 &\  0 &\  0 \\
  1 &\  0 &\  0 \\
\end{array}%
\right],\ \
M_y=\frac{1}{\sqrt{2}}\left[%
\begin{array}{ccc}
  0 &\  -i &\  -i \\
  i &\  0 &\  0 \\
  i &\  0 &\  0 \\
\end{array}%
\right],\ \
M_z=\left[%
\begin{array}{ccc}
  0 &\  0 &\  0 \\
  0 &\  1 &\  0 \\
  0 &\  0 &\  -1 \\
\end{array}%
\right],\ \
\end{eqnarray}
are spin 1 operators, obeying the
commutation relation $[M_p,M_q]=i\varepsilon^{pqr}M_r$ with the
Levi-Civita symbol $\varepsilon^{pqr}$.

As shown in Ref. \cite{BaksicPRL116}, moving to ``dressed-state
picture'', one can chose a picture transformation
\begin{eqnarray}\label{e15}
V(t)=\exp[i\eta(t)M_z]\exp[i\mu(t)M_x]\exp[i\xi(t)M_z],
\end{eqnarray}
which is parametrized as a rotation of spin with Euler angles
$\xi(t)$, $\mu(t)$, and $\eta(t)$. Moreover, to full fill the
condition $V(0)=V(T)=1$, the angle $\mu(t)$ should satisfy
$\mu(0)=\mu(T)=0(2\pi)$, and the other two angles can have arbitrary
values. If we want the correction Hamiltonian $H_{co}(t)$ to has the
same form as $H_{eff}(t)$, $H_{co}(t)$ can be chosen to have the
general form
\begin{eqnarray}\label{e16}
H_{co}(t)=U(t)(g_x(t)M_x+g_z(t)M_z)U^{\dag}(t),
\end{eqnarray}
where $g_x(t)$ and $g_z(t)$ are two controlled parameters.
Therefore, what we need is only a simple modification of the
original angle $\theta(t)$ and amplitude $\Omega(t)$ as
\begin{eqnarray}\label{e17}
&&\theta(t)\rightarrow\tilde{\theta}(t)=\theta(t)-\arctan(\frac{g_x(t)}{\Omega(t)+g_z(t)}),\cr\cr&&
\Omega(t)\rightarrow\tilde{\Omega}(t)=\sqrt{(\Omega(t)+g_z(t))^2+g_x^2(t)}.
\end{eqnarray}

In addition, to cancel the unwanted transitions between dressed
states in the ``dressed-state picture'', the controlled parameters
should be chosen as
\begin{eqnarray}\label{e14.1}
&&g_x(t)=\frac{\dot{\mu}}{\cos\xi}-\dot{\theta}\tan\xi,\cr\cr&&
g_z(t)=-\Omega+\dot{\xi}+\frac{\dot{\mu}\sin\xi-\dot{\theta}}{\tan\mu\cos\xi},
\end{eqnarray}
and they are independent of $\eta(t)$. Moreover, the population of
the intermediate state $|\phi_0\rangle$ is given by
\begin{eqnarray}\label{e15.1}
|\langle\Psi(t)|\phi_0(t)\rangle|=\sin^2\mu(t)\cos^2\xi(t).
\end{eqnarray}
For simplicity, we choose $\xi(t)\equiv0$. To full fill boundary
condition $\mu(0)=\mu(T)=0(2\pi)$, $\theta(0)=0$ and
$\theta(T)=\pi/2$ as well as avoid the singularity of the expression
for each pulse, we adopt the following parameters
\begin{eqnarray}\label{e16.1}
&&\theta(t)=\frac{\pi t}{2T}-\frac{1}{3}\sin(\frac{2\pi
t}{T})+\frac{1}{24}\sin(\frac{4\pi t}{T}),\cr\cr&&
\dot{\theta}(t)=\frac{4\pi}{3T}\sin^4(\frac{\pi t}{T}),\cr\cr&&
\mu(t)=\frac{A}{2}[1-\cos(\frac{2\pi t}{T})],\cr\cr&&
\dot{\mu}(t)=\frac{\pi A}{T}\sin(\frac{2\pi t}{T}),
\end{eqnarray}
where $A$ is a time-independent parameter which controls the maximal
value of $\mu(t)$. If we set $0<A<\pi/2$, when $A$ decreases, the
population of intermediate state $|\phi_0\rangle$ also decreases,
however, according to the expression of $g_z(t)$, the value of
$\tilde{\Omega}(t)\times T$ will increase; that means one has to
increase the interaction time $T$ when the pulses' amplitudes
$\tilde{\Omega}(t)$ has a fixed value. Therefore, it is better to
choose a suitable $A$, so that both population of intermediate state
$|\phi_0\rangle$ and interaction time can be restricted in a desired
range. We find that $A=0.5$ can meet our requirement, which gives
$|\langle\Psi(t)|\phi_0(t)\rangle|=\sin^2\mu(t)\leq0.23$ and
$\tilde{\Omega}(t)\times T\approx7$. Till now, there is still a
question being remained, that is, the expressions of pulses
$\tilde{\Omega}_a(t)=\tilde{\Omega}(t)\cos\tilde{\theta}(t)$ and
$\tilde{\Omega}_b(t)=\tilde{\Omega}(t)\sin\tilde{\theta}(t)$ are too
complex for realization in experiments. In order to make the
protocol more feasible in experiments, the Rabi frequencies of
pulses should be expressed by some frequently used functions, e.g.
Gaussian functions and sine functions, or their linear
superpositions. Thanks to the curve fitting, we find two replaceable
pulses $\bar{\Omega}_a(t)$ and $\bar{\Omega}_b(t)$ respectively for
$\tilde{\Omega}_a(t)$ and $\tilde{\Omega}_b(t)$ as
\begin{eqnarray}\label{e17.1}
&&\bar{\Omega}_a(t)=\zeta_{a_1}e^{-[(t-\tau_{a_1})/\chi_{a_1}]^2}+\zeta_{a_2}e^{-[(t-\tau_{a_2})/\chi_{a_2}]^2},\cr\cr&&
\bar{\Omega}_b(t)=\zeta_{b_1}e^{-[(t-\tau_{b_1})/\chi_{b_1}]^2}+\zeta_{b_2}e^{-[(t-\tau_{b_2})/\chi_{b_2}]^2},
\end{eqnarray}
where,
\begin{eqnarray}\label{e18}
&&\zeta_{a_1}=6.226/T,\ \zeta_{a_2}=1.332/T,\ \zeta_{b_1}=6.226/T,\
\zeta_{b_2}=1.332/T,\cr\cr&& \tau_{a_1}=0.597T,\
\tau_{a_2}=0.2395T,\ \tau_{b_1}=0.4033T,\
\tau_{b_2}=0.7605T,\cr\cr&& \chi_{a_1}=0.2214T,\
\chi_{a_2}=0.1971T,\ \chi_{b_1}=0.2214T,\ \chi_{b_2}=0.1971T.
\end{eqnarray}
Here, $\zeta_{\alpha_\beta}$ ($\alpha=a,b$, $\beta=1,2$) is the
pulse amplitude of the $\beta$-th component in pulse
$\Omega_\alpha(t)$, $\tau_{\alpha_\beta}$ describes the extreme
point of the $\beta$-th component in pulse $\Omega_\alpha(t)$, and
the $\chi_{\alpha_\beta}$ controls the width of $\beta$-th component
in pulse $\Omega_\alpha(t)$.
\begin{figure}
\scalebox{0.6}{\includegraphics[scale=1]{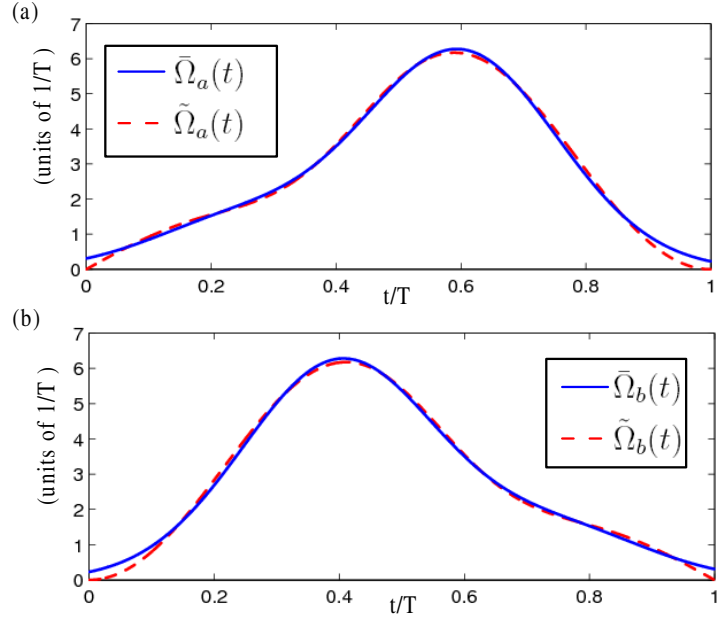}} \caption{(a)
Comparison between $\tilde{\Omega}_a(t)$ (the dashed red line) and
$\bar{\Omega}_a(t)$ (the solid blue line) (versus $t/T$). (b)
Comparison between $\tilde{\Omega}_b(t)$ (the dashed red line) and
$\bar{\Omega}_b(t)$ (the solid blue line) (versus $t/T$).}
\end{figure}
As a comparison, we plot $\tilde{\Omega}_a(t)$
($\tilde{\Omega}_b(t)$) with $\bar{\Omega}_a(t)$
($\bar{\Omega}_b(t)$) versus $t/T$ in Fig. 2 (a) (Fig. 2 (b)). As
shown in Fig. 2, the solid blue curve for $\bar{\Omega}_a(t)$
($\bar{\Omega}_b(t)$) and the dashed red curve for
$\tilde{\Omega}_a(t)$ ($\tilde{\Omega}_b(t)$) are considerably close
to each other. In the next section, pulses with Rabi frequencies
$\bar{\Omega}_1(t)=\sqrt{2}\bar{\Omega}_a(t)$,
$\bar{\Omega}_2(t)=\sqrt{2}\bar{\Omega}_a(t)$,
$\bar{\Omega}_3(t)=\sqrt{2}\bar{\Omega}_a(t)$ and
$\bar{\Omega}_4(t)=\sqrt{2}\bar{\Omega}_b(t)$ will be demonstrated
to drive the system from its initial state
$|\Psi(0)\rangle=|\psi_1\rangle$ to the target state
$|\Psi(T)\rangle=|W\rangle$ with high fidelity via numerical
simulations for the sake of proving the replacements here for the
Rabi frequencies of the pulses are effective.

\section{Numerical simulations}

In this section, we will investigate the performance of the protocol
via numerical simulations. The fidelity of the target state
$|W\rangle$ is defined as $F(t)=|\langle W|\rho(t)|W\rangle|$, where
$\rho(t)$ is the density operator of the system. Firstly, as
condition $\Omega_a,\Omega_b\ll g$ is set to obtain the effective
Hamiltonian $H_{eff}(t)$, so before doing numerical simulations and
further discussions based on the original Hamiltonian $H_I(t)$ in
the interaction picture, we need to choose a suitable value for
coupling constant $g$.
\begin{figure}
\scalebox{0.6}{\includegraphics[scale=1]{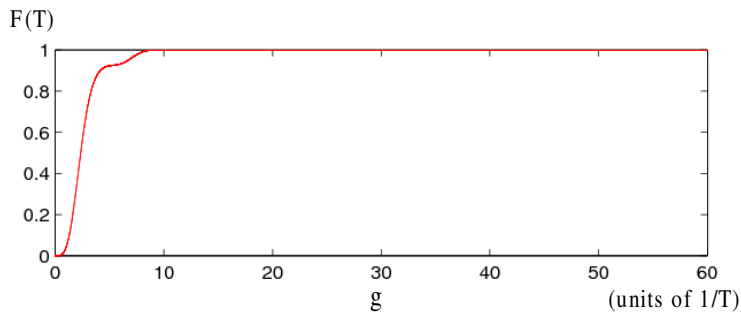}} \caption{The
final fidelity $F(T)$ versus $g$.}
\end{figure}
In present protocol, the pulses' amplitudes are
$\bar{\Omega}_0=\max\limits_{\mathop{0\leq t\leq
T}\limits_{k=1,2,3,4}}\{\bar{\Omega}_k(t)\}\approx9.8/T$, and
condition $\Omega_a,\Omega_b\ll g$ can be replaced by
$\bar{\Omega}_0\ll g$. Seen from Fig. 3, the final fidelity $F(T)$
is almost 1 when $g\geq10/T$. That means even if condition
$\bar{\Omega}_0\ll g$ is violated, one can also obtain a W state by
using the present protocol. Generally speaking, since the coupling
constant $g$ has an upper limit in real experiments, the condition
$\bar{\Omega}_0\ll g$ may cause the speed limit of the system's
evolution. But when $\bar{\Omega}_0\ll g$ is full filled, the system
is guided by the effective Hamiltonian $H_{eff}(t)$, so the dark
state $|\phi_0\rangle$ of $H_c$ has an absolutely predominance among
all the intermediate states. Since $|\phi_0\rangle$ has a lower
energy compared with other eigenstates of $H_c$, using
$|\phi_0\rangle$ as the intermediate state while restraining
populations for other eigenstates of $H_c$ can help us to reduce the
dissipation. However, when $g$ is not large enough, the system will
evolve along an unknown path, which does not decided by the
effective Hamiltonian. As a result, the population of each
intermediate state can not be forecasted as before, meanwhile
$|\phi_0\rangle$ does not predominant in this case. Thus dissipation
will increase, finally resulting in a relatively bad performance
when decoherence mechanisms are taken into account. Therefore, for
both high speed and robustness against dissipation, we adopt
$g=30/T$, slightly larger than $\bar{\Omega}_0$
($\bar{\Omega}_0/g\approx0.33$). After coupling constant $g$ being
chosen, we would like to examine the population
$P_\iota=\langle\psi_\iota|\rho(t)|\psi_\iota\rangle$
($\iota=1,2,\cdots,9$) of state $|\psi_\iota\rangle$ during the
evolution. So, we plot $P_\iota$ versus $t/T$ in Fig. 4.
\begin{figure}
\scalebox{0.6}{\includegraphics[scale=1]{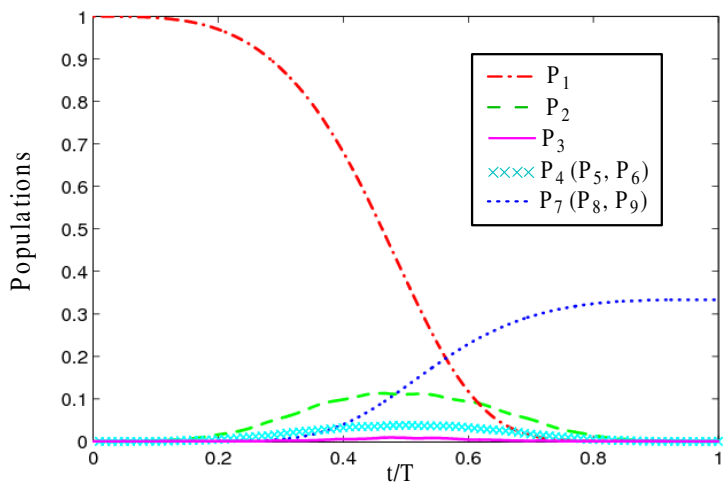}} \caption{The
population $P_\iota$ \textbf{($\iota=1, 2, 3, \cdots, 9$)} of state
$|\psi_\iota\rangle$ versus $t/T$. $P_1$: the dashed and dotted red
line. $P_2$: the dashed green line. $P_3$: the solid pink line.
$P_4$, $P_5$, $P_6$: the light blue crosses. $P_7$, $P_8$ and $P_9$:
the dotted blue line.}
\end{figure}
As shown in Fig. 4, the population $|\psi_3\rangle$ (see the solid
pink line of Fig. 4), which is not the component of
$|\phi_0\rangle$, keeps nearly 0 during the evolution. This result
is coincide with the dynamics governed by $H_{eff}(t)$. Finally at
$t=T$, the target state $|W\rangle$ can be obtained.

Secondly, since accelerating the adiabatic passage is a purpose for
implementing the present protocol, it is necessary to show the
present protocol is faster than preparing W state with adiabatic
passages. Here, considering STIRAP is a famous method for the
adiabatic passages, we start with constructing an adiabatic passage
to prepare W state by using STIRAP. We can design the Rabi
frequencies of pulses as
\begin{eqnarray}\label{e19}
\Omega_1'(t)=\Omega_2'(t)=\Omega_3'(t)=\Omega_0'e^{-[(t-t_0-T/2)/t_c]^2},\
\Omega_4'(t)=\Omega_0'e^{-[(t+t_0-T/2)/t_c]^2},
\end{eqnarray}
where, $\Omega_0'$ is the pulses' amplitudes for STIRAP, $t_0=0.15T$
and $t_c=0.2T$ are two related parameters. Then, to compare the
present protocol with that by STIRAP, we plot Fig. 5 to show the
fidelities of obtaining the target state $|W\rangle$ versus $t/T$
with different methods.
\begin{figure}
\scalebox{0.6}{\includegraphics[scale=1]{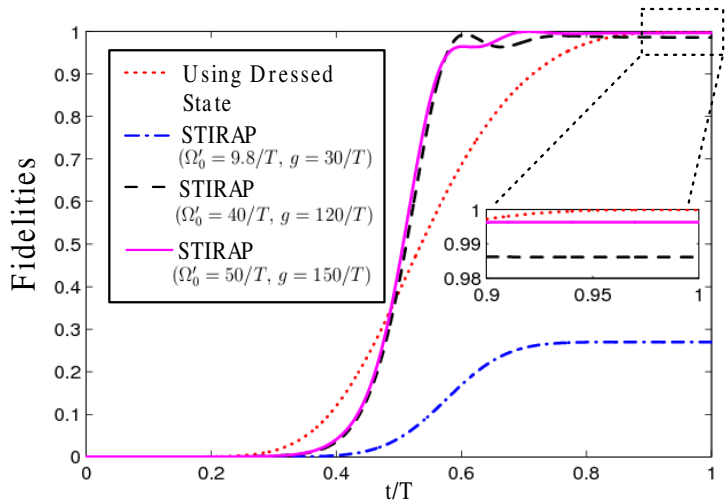}} \caption{The
fidelities of the target state $|W\rangle$ versus $t/T$ with
different methods. The dotted red line: using dressed method. The
dashed and dotted blue line: STIRAP with $\Omega_0=9.8/T$ and
$g=30/T$. The dashed black line: STIRAP with $\Omega_0=40/T$ and
$g=120/T$. The solid pink line: STIRAP with $\Omega_0=50/T$ and
$g=150/T$.}
\end{figure}
As shown in Fig. 5, the fidelity of the present protocol can reach 1
at $t=T$ (see the dotted red line in Fig. 5) while with the same
condition for STIRAP ($\Omega_0'=9.8/T$, $g=30/T$, see the dashed
and dotted blue line in Fig. 5), the fidelity is only about 0.275
due to the badly violation of the adiabatic condition. So we
increase the pulses' amplitudes $\Omega_0'$ and the coupling
constant $g$ to $40/T$ and $120/T$, respectively. In this case (see
the dashed black line in Fig. 5), the fidelity can increase close to
1, however, the final fidelity is only 0.985, still a little
disappointing. Finally, when the pulses' amplitudes $\Omega_0'$ are
increased to $50/T$, and $g$ is increased to $150/T$, the fidelity
even more approach to 1 (above 0.99), however its performance is
still worse than that of the present protocol (see the solid pink
line in Fig. 5). As we mentioned in Sec. III, for a relatively high
speed, the product of the pulses' amplitudes $\Omega_0$ and the
total interaction time $T$ is the smaller the better. Because when
$\Omega_0$ takes a fixed value (such as the upper limit for the
system), the one has the smaller product $\Omega_0\times T$ will
have less interaction time. In the present protocol, the pulses'
amplitudes $\bar{\Omega}_0$ is only $9.8/T$, while for STIRAP, to
obtain an enough high fidelity, one should set $\Omega_0'\geq50/T$.
Therefore, the speed of the present protocol to obtain the target
state is faster a lot compared with that with STIRAP.

\begin{figure}
\scalebox{0.4}{\includegraphics[scale=1]{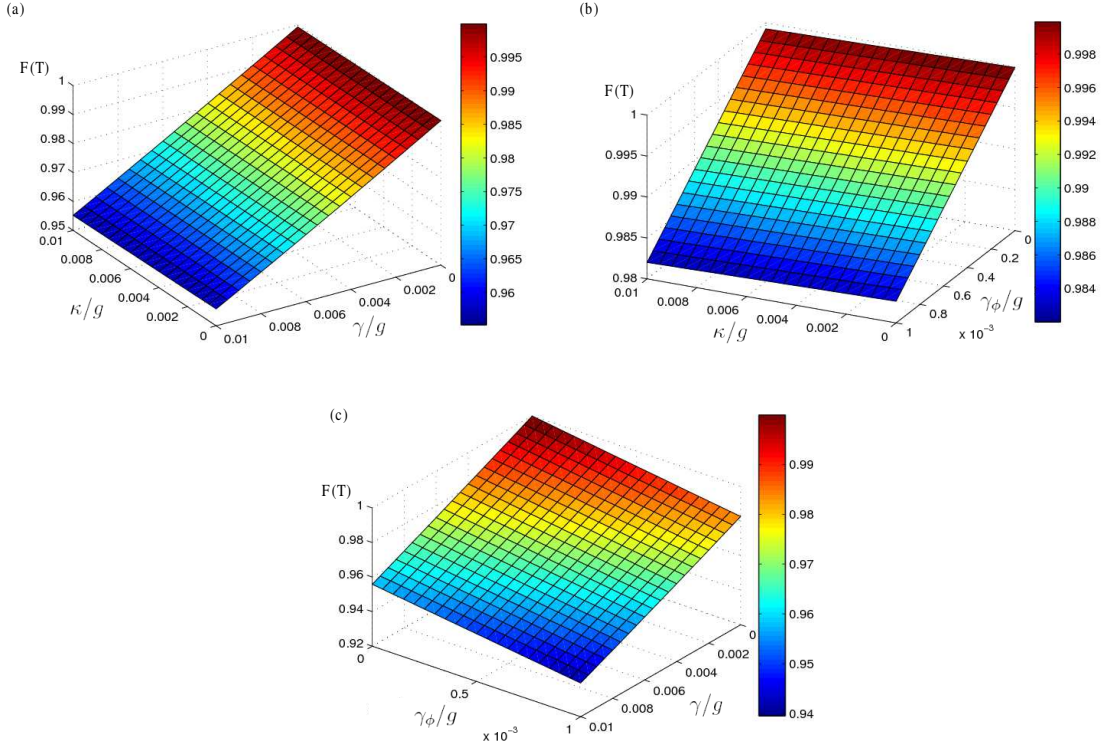}} \caption{(a) The
final fidelity $F(T)$ versus $\kappa/g$ and $\gamma/g$. (b) The
final fidelity $F(T)$ versus $\kappa/g$ and $\gamma_\phi/g$. (c) The
final fidelity $F(T)$ versus $\gamma/g$ and $\gamma_\phi/g$.}
\end{figure}
Thirdly, in real experiments, the dissipation caused by decoherence
mechanisms are ineluctable. Therefore, we would like to check the
fidelity $F(T)$ when decoherence mechanisms are taken into account
in order to help us to forecast the experimental feasibility. In the
present protocol, the major factors of decoherence mechanisms are
(i) cavity decay (with decay rate $\kappa$), (ii) the spontaneous
emissions from $|e\rangle_k$ to $|0\rangle_k$ and $|1\rangle_k$ with
spontaneous emission rates $\gamma_{0k}$ and $\gamma_{1k}$,
respectively, (iii) the dephasing between $|e\rangle_k$ and
$|0\rangle_k$ ($|e\rangle_k$ and $|1\rangle_k$) with dephasing rate
$\gamma_{\phi0k}$ ($\gamma_{\phi1k}$) ($k=1,2,3,4$). The evolution
of the system can be described by a master equation in Lindblad form
as following
\begin{equation}\label{e20}
\dot{\rho}=i[\rho,H_I]+\sum\limits_{l}[L_l\rho
L_l^{\dagger}-\frac{1}{2}(L_l^{\dagger}L_l\rho+\rho
L_l^{\dagger}L_l)],
\end{equation}
where, $L_l$ ($l=1,2,3,\cdots,17$) is the Lindblad operator. There
are seventeen Lindblad operators in the present protocol as
\begin{eqnarray}\label{e21}
&&L_{1}=\sqrt{\gamma_{11}}|1\rangle_1\langle e|,\ \ \
L_{2}=\sqrt{\gamma_{12}}|1\rangle_2\langle e|,\ \ \
L_{3}=\sqrt{\gamma_{13}}|1\rangle_3\langle e|,\ \ \
L_{4}=\sqrt{\gamma_{14}}|1\rangle_4\langle e|,\cr\cr&&
L_{5}=\sqrt{\gamma_{01}}|0\rangle_1\langle e|,\ \ \
L_{6}=\sqrt{\gamma_{02}}|0\rangle_2\langle e|,\ \ \
L_{7}=\sqrt{\gamma_{03}}|0\rangle_3\langle e|,\ \ \
L_{8}=\sqrt{\gamma_{04}}|0\rangle_4\langle e|,\cr\cr&&
L_{9}=\sqrt{\gamma_{\phi11}/2}(|e\rangle_1\langle
e|-|1\rangle_1\langle1|),\ \ \
L_{10}=\sqrt{\gamma_{\phi12}/2}(|e\rangle_2\langle
e|-|1\rangle_2\langle1|),\cr\cr&&
L_{11}=\sqrt{\gamma_{\phi13}/2}(|e\rangle_3\langle
e|-|1\rangle_3\langle1|),\ \ \ L_{12}=\sqrt{\gamma_{\phi
14}/2}(|e\rangle_4\langle e|-|1\rangle_4\langle1|),\cr\cr&&
L_{13}=\sqrt{\gamma_{\phi01}/2}(|e\rangle_1\langle
e|-|0\rangle_1\langle0|),\ \ \
L_{14}=\sqrt{\gamma_{\phi02}/2}(|e\rangle_2\langle
e|-|0\rangle_2\langle0|),\cr\cr&&
L_{15}=\sqrt{\gamma_{\phi03}/2}(|e\rangle_3\langle
e|-|0\rangle_3\langle0|),\ \ \
L_{16}=\sqrt{\gamma_{\phi04}/2}(|e\rangle_4\langle
e|-|0\rangle_4\langle0|),\cr\cr&& L_{17}=\sqrt{\kappa}a.
\end{eqnarray}
For simplicity, we assume $\gamma_{1k}=\gamma_{0k}=\gamma$ and
$\gamma_{\phi1k}=\gamma_{\phi0k}=\gamma_{\phi}$ in the following
discussions. The final fidelity $F(T)$ versus $\kappa/g$ and
$\gamma/g$ is given in Fig. 6 (a), the final fidelity $F(T)$ versus
$\kappa/g$ and $\gamma_{\phi}/g$ is given in Fig. 6 (b), and the
final fidelity $F(T)$ versus $\gamma/g$ and $\gamma_\phi/g$ is given
in Fig. 6 (c). Some samples of the final fidelity $F(T)$ with
corresponding $\kappa/g$, $\gamma/g$ and $\gamma_{\phi}/g$ are given
in Table I.
\begin{center}{\bf Table I. Samples of the final fidelity $F(T)$ with corresponding $\kappa/g$, $\gamma/g$ and $\gamma_{\phi}/g$.\ \ \ \ \ \ \ \ \ \ \ \
\ \ \ \ \ \ \ \ \ \ \  }{\small
\begin{tabular}{cccc}\hline\hline
$\kappa/g\ (\times10^{-2})$ \ \ \ \ \ \ \ \ \ \ \ \ \ \ \ \ \ \ \ \ \ \ \ &$\gamma/g\ (\times10^{-2})$ \ \ \ \ \ \ \ \ \ \ \ \ \ \ \ \ \ \ \ \ \ &$\gamma_{\phi}/g\ (\times10^{-3})$ \ \ \ \ \ \ \ \ \ \ \ \ \ \ \ \ \ \ \ \ \ \ \ &$F(T)$\\
\hline $\ \ 1 \ \ \ \ \  \ \ \ \ \ \ \ \ \ \ \ \ \ \ \ \ \ \ \ $&$1\ \ \ \ \ \ \ \ \ \ \ \ \ \ \ \ \ \ \ \ \ $&$1\ \ \ \ \ \ \ \ \ \ \ \ \ \ \ \ \ \ \ $&$0.9389$\\
$\ \ 1 \ \ \ \ \ \ \ \ \ \ \ \ \ \ \ \ \ \ \ \ \ \ \ \ $&$1\ \ \ \ \ \ \ \ \ \ \ \ \ \ \ \ \ \ \ \ \ $&$0.8 \ \ \ \ \ \ \ \ \ \ \ \ \ \ \ \ \ \ \ $&$0.9421$\\
$1 \ \ \ \ \ \ \ \ \ \ \ \ \ \ \ \ \ \ \ \ \ \ $&$0.8\ \ \ \ \ \ \ \ \ \ \ \ \ \ \ \ \ \ \ \ \ $&$ 1\ \ \ \ \ \ \ \ \ \ \ \ \ \ \ \ \ \ \ $&$0.9473$\\
$0.8 \ \ \ \ \ \ \ \ \ \ \ \ \ \ \ \ \ \ \ \ \ $&$1\ \ \ \ \ \ \ \ \ \ \ \ \ \ \ \ \ \ \ \ \ $&$ 1 \ \ \ \ \ \ \ \ \ \ \ \ \ \ \ \ \ \ \ $&$0.9390$\\
$0.8 \ \ \ \ \ \ \ \ \ \ \ \ \ \ \ \ \ \ \ \ \ $&$0.8\ \ \ \ \ \ \ \ \ \ \ \ \ \ \ \ \ \ \ \ \ $&$ 0.8 \ \ \ \ \ \ \ \ \ \ \ \ \ \ \ \ \ \ \ $&$0.9507$\\
$0.8 \ \ \ \ \ \ \ \ \ \ \ \ \ \ \ \ \ \ \ \ \ $&$0.8\ \ \ \ \ \ \ \ \ \ \ \ \ \ \ \ \ \ \ \ \ $&$ 0.5 \ \ \ \ \ \ \ \ \ \ \ \ \ \ \ \ \ \ \ $&$0.9556$\\
$0.8 \ \ \ \ \ \ \ \ \ \ \ \ \ \ \ \ \ \ \ \ \ $&$0.5\ \ \ \ \ \ \ \ \ \ \ \ \ \ \ \ \ \ \ \ \ $&$ 0.8 \ \ \ \ \ \ \ \ \ \ \ \ \ \ \ \ \ \ \ $&$0.9635$\\
$0.5 \ \ \ \ \ \ \ \ \ \ \ \ \ \ \ \ \ \ \ \ \ $&$0.8\ \ \ \ \ \ \ \ \ \ \ \ \ \ \ \ \ \ \ \ \ $&$ 0.8\ \ \ \ \ \ \ \ \ \ \ \ \ \ \ \ \ \ \ $&$0.9509$\\
$0.5 \ \ \ \ \ \ \ \ \ \ \ \ \ \ \ \ \ \ \ \ \ $&$0.5\ \ \ \ \ \ \ \ \ \ \ \ \ \ \ \ \ \ \ \ \ $&$ 0.5\ \ \ \ \ \ \ \ \ \ \ \ \ \ \ \ \ \ \ $&$0.9687$\\
$0.5 \ \ \ \ \ \ \ \ \ \ \ \ \ \ \ \ \ \ \ \ \ $&$0.5\ \ \ \ \ \ \ \ \ \ \ \ \ \ \ \ \ \ \ \ \ $&$ 0.3\ \ \ \ \ \ \ \ \ \ \ \ \ \ \ \ \ \ \ $&$0.9721$\\
$0.5 \ \ \ \ \ \ \ \ \ \ \ \ \ \ \ \ \ \ \ \ \ $&$0.3\ \ \ \ \ \ \ \ \ \ \ \ \ \ \ \ \ \ \ \ \ $&$ 0.5\ \ \ \ \ \ \ \ \ \ \ \ \ \ \ \ \ \ \ $&$0.9775$\\
$0.3 \ \ \ \ \ \ \ \ \ \ \ \ \ \ \ \ \ \ \ \ \ $&$0.5\ \ \ \ \ \ \ \ \ \ \ \ \ \ \ \ \ \ \ \ \ $&$ 0.5\ \ \ \ \ \ \ \ \ \ \ \ \ \ \ \ \ \ \ $&$0.9659$\\
$0.3 \ \ \ \ \ \ \ \ \ \ \ \ \ \ \ \ \ \ \ \ \ $&$0.3\ \ \ \ \ \ \ \ \ \ \ \ \ \ \ \ \ \ \ \ \ $&$ 0.3 \ \ \ \ \ \ \ \ \ \ \ \ \ \ \ \ \ \ \ $&$0.9811$\\
$0.3 \ \ \ \ \ \ \ \ \ \ \ \ \ \ \ \ \ \ \ \ \ $&$0.3\ \ \ \ \ \ \ \ \ \ \ \ \ \ \ \ \ \ \ \ \ $&$ 0.1 \ \ \ \ \ \ \ \ \ \ \ \ \ \ \ \ \ \ \ $&$0.9845$\\
$0.3 \ \ \ \ \ \ \ \ \ \ \ \ \ \ \ \ \ \ \ \ \ $&$0.1\ \ \ \ \ \ \ \ \ \ \ \ \ \ \ \ \ \ \ \ \ $&$ 0.3 \ \ \ \ \ \ \ \ \ \ \ \ \ \ \ \ \ \ \ $&$0.9900$\\
$0.1 \ \ \ \ \ \ \ \ \ \ \ \ \ \ \ \ \ \ \ \ \ $&$0.3\ \ \ \ \ \ \ \ \ \ \ \ \ \ \ \ \ \ \ \ \ $&$0.3\ \ \ \ \ \ \ \ \ \ \ \ \ \ \ \ \ \ \ $&$0.9812$\\
$0.1 \ \ \ \ \ \ \ \ \ \ \ \ \ \ \ \ \ \ \ \ \ $&$0.1\ \ \ \ \ \ \ \ \ \ \ \ \ \ \ \ \ \ \ \ \ $&$0.1\ \ \ \ \ \ \ \ \ \ \ \ \ \ \ \ \ \ \ $&$0.9936$\\
\hline \hline
\end{tabular}}
\end{center}
According to Fig. 6 and Table I, we have the following results. (i)
$F(T)$ is very robust against the cavity decay since the population
of $|\psi_3\rangle$ is restrained (see Fig. 4). (ii) $F(T)$ is more
sensitive to the spontaneous emissions than the cavity decay.
However, when $\gamma/g$ increases from 0 to 0.01, $F(t)$ keeps
higher than 0.957 with $\gamma_\phi=0$ and $\kappa=0$. We can say
the present protocol to prepare W states is also robust against the
spontaneous emissions. (iii) The dephasing influences $F(T)$ mostly.
When $\gamma_\phi/g$ increases from 0 to only $1\times10^{-3}$,
$F(T)$ falls from 1 to 0.983. We also investigate the performance of
STIRAP when dephasing is taken into account. As a comparison, we
plot the final fidelities $F(T)$ versus $\gamma_\phi/g$ for both
present protocol (the dashed red line) and STIRAP protocol (the
solid blue line) in Fig. 7.
\begin{figure}
\scalebox{0.6}{\includegraphics[scale=1]{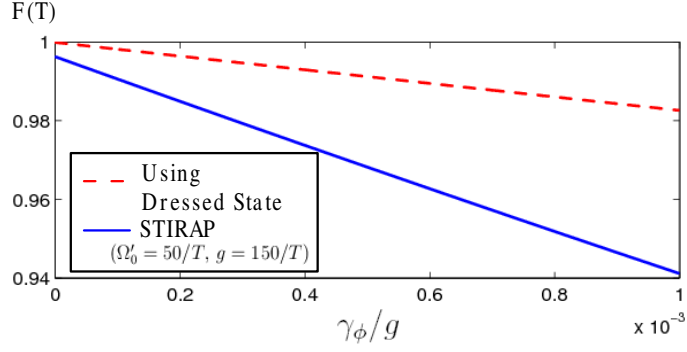}} \caption{The
final fidelities $F(T)$ versus $\gamma_\phi/g$ for the present
protocol (the dashed red line) and that with STIRAP (the solid blue
line).}
\end{figure}
As shown in Fig. 7, with STIRAP, $F(T)$ decreases from 1 to 0.942
when $\gamma_\phi/g$ increases from 0 to $1\times10^{-3}$. Comparing
with STIRAP, it is obvious that the present protocol is more robust
against dephasing on account of the acceleration for the evolution
speed. In addition, Refs. \cite{YangPRL92,XiangRMP85} have shown
that $g\sim180$MHz, $\gamma\sim1.32$MHz, $\kappa\sim1.32$MHz,
$\gamma_\phi\sim10$kHz can be realized in real experiments.
Submitting these parameters into Eq. (\ref{e20}) and Eq.
(\ref{e21}), we have $F(T)=0.9659$. Therefore, the present protocol
could work well when decoherence mechanisms are considered.

Fourthly, due to the variations of the parameters caused by the
experimental imperfection operations, the evolution of the system
will deviate from our expectation. It is worthwhile to investigate
the influences from variations of the parameters caused by the
experimental imperfection. Here we would like to discuss the
variations $\delta T$, $\delta\bar{\Omega}_0$ and $\delta g$ of the
total evolution time $T$, pulses' amplitudes $\bar{\Omega}_0$ and
the coupling constant $g$, respectively.
\begin{figure}
\scalebox{0.4}{\includegraphics[scale=1]{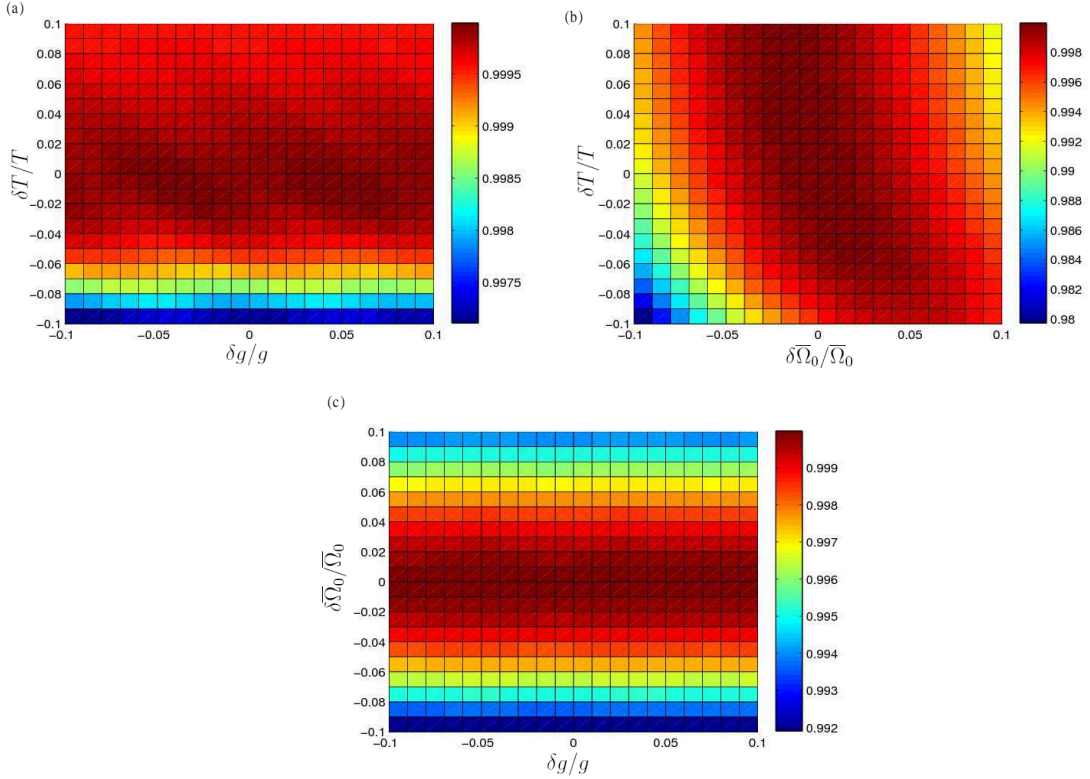}} \caption{(a) The
final fidelity $F(T')$ versus $\delta T/T$ and $\delta g/g$. (b) The
final fidelity $F(T')$ versus $\delta T/T$ and
$\delta\bar{\Omega}_0/\bar{\Omega}_0$. (c) The final fidelity $F(T)$
versus $\delta\overline{\Omega}_0/\overline{\Omega}_0$ and $\delta
g/g$.}
\end{figure}
We assume that $T'=T+\delta T$ is the erroneous total interaction
time when there is a variation $\delta T$ for the original
interaction time. We plot $F(T')$ versus $\delta T/T$ and $\delta
g/g$ in Fig. 8 (a), $F(T')$ versus $\delta T/T$ and
$\delta\bar{\Omega}_0/\bar{\Omega}_0$ in Fig. 8 (b), and $F(T)$
versus $\delta g/g$ and $\delta\bar{\Omega}_0/\bar{\Omega}_0$ in
Fig. 8 (c). Some samples of the final fidelity $F(T')$ with
corresponding $\delta T/T$, $\delta\bar{\Omega}_0/\bar{\Omega}_0$
and $\delta g/g$ are given in Table II. According to Fig. 8 and
Table. II, we can obtain following results. (i) Seen from Fig. 8
(a), $F(T')$ is insensitive to the variation $\delta g$ for the
coupling strength. Besides, Fig. 8 (c) shows that $F(T)$ is almost
not influenced by the variation $\delta g$. This result is because
we have chosen a suitable coupling constant $g=30/T$ in the first
part of discussions. It is also shown in Fig. 3 that, the final
fidelity is nearly 1 when $g\geq10/T$. Therefore, the coupling
constant we chosen is good enough to resist the variation $\delta
g$. (ii) As shown in Figs. 8 (a) and (b), $F(T')$ is also very
robust to the variation $\delta T$ of the total interaction time.
When $T'=0.9T$ with $\delta g=0$ and $\delta\bar{\Omega}_0=0$, the
fidelity only decreases about 0.003. Moreover, when $T'>T$, the
fidelity is almost unchange and close to 1, on account of the
suitable boundary condition for controlled parameters (e.g.
$\theta$, $\dot{\theta}$, $\mu$ and $\dot{\mu}$) set in Sec. III.
(iii) The variation $\delta\bar{\Omega}_0$ of pulses' amplitudes
$\bar{\Omega}_0$ influences the fidelity mostly according to Figs. 8
(b) and (c). However, as shown in Fig. 8 (b), $F(T')$ is still
higher than 0.98 even when
$|\delta\bar{\Omega}_0/\bar{\Omega}_0|=|\delta T/T|=10\%$, and as
shown in Fig. 8 (c), $F(T')$ is still higher than 0.992 even when
$|\delta\bar{\Omega}_0/\bar{\Omega}_0|=|\delta g/g|=10\%$. This
indicates that the present protocol holds robustness against the
variation $\delta\bar{\Omega}_0$ as well. (iv) There is an
interesting phenomenon shown in Fig. 8 (b), i.e., when
$\delta\bar{\Omega}_0$ and $\delta T$ have the same sign (both
positive or both negative), the fidelity still keeps in a high
level. This tells us that, if we have the smaller (larger) pulses'
amplitudes than the designed one, we should increase (reduce)
interaction time to correct the error. Based on the discussions
above, we conclude that the present protocol is robust against the
variations $\delta T$, $\delta\bar{\Omega}_0$ and $\delta g$.
\begin{center}{\bf Table II. Samples of the final fidelity $F(T')$ with corresponding $\delta T/T$, $\delta\bar{\Omega}_0/\bar{\Omega}_0$ and $\delta g/g$.\ \ \ \ \ \ \ \ \ \ \ \
\ \ \ \ \ \ \ \ \ \ \  }{\small
\begin{tabular}{cccc}\hline\hline
$\delta T/T$ \ \ \ \ \ \ \ \ \ \ \ \ \ \ \ \ \ \ \ \ \ \ \ &$\delta\bar{\Omega}_0/\bar{\Omega}_0$ \ \ \ \ \ \ \ \ \ \ \ \ \ \ \ \ \ \ \ \ \ &$\delta g/g$ \ \ \ \ \ \ \ \ \ \ \ \ \ \ \ \ \ \ \ \ \ \ \ &$F(T')$\\
\hline $\ \ 10\% \ \ \ \ \  \ \ \ \ \ \ \ \ \ \ \ \ \ \ \ \ \ \ \ $&$10\%\ \ \ \ \ \ \ \ \ \ \ \ \ \ \ \ \ \ \ \ \ $&$10\%\ \ \ \ \ \ \ \ \ \ \ \ \ \ \ \ \ \ \ $&$0.9907$\\
$\ \ 10\% \ \ \ \ \ \ \ \ \ \ \ \ \ \ \ \ \ \ \ \ \ \ \ \ $&$10\%\ \ \ \ \ \ \ \ \ \ \ \ \ \ \ \ \ \ \ \ \ $&$-10\% \ \ \ \ \ \ \ \ \ \ \ \ \ \ \ \ \ \ \ $&$0.9907$\\
$10\% \ \ \ \ \ \ \ \ \ \ \ \ \ \ \ \ \ \ \ \ \ \ $&$-10\%\ \ \ \ \ \ \ \ \ \ \ \ \ \ \ \ \ \ \ \ \ $&$ 10\%\ \ \ \ \ \ \ \ \ \ \ \ \ \ \ \ \ \ \ $&$0.9944$\\
$10\% \ \ \ \ \ \ \ \ \ \ \ \ \ \ \ \ \ \ \ \ \ $&$-10\%\ \ \ \ \ \ \ \ \ \ \ \ \ \ \ \ \ \ \ \ \ $&$ -10\% \ \ \ \ \ \ \ \ \ \ \ \ \ \ \ \ \ \ \ $&$0.9944$\\
$-10\% \ \ \ \ \ \ \ \ \ \ \ \ \ \ \ \ \ \ \ \ \ $&$10\%\ \ \ \ \ \ \ \ \ \ \ \ \ \ \ \ \ \ \ \ \ $&$ 10\% \ \ \ \ \ \ \ \ \ \ \ \ \ \ \ \ \ \ \ $&$0.9965$\\
$-10\% \ \ \ \ \ \ \ \ \ \ \ \ \ \ \ \ \ \ \ \ \ $&$10\%\ \ \ \ \ \ \ \ \ \ \ \ \ \ \ \ \ \ \ \ \ $&$ -10\% \ \ \ \ \ \ \ \ \ \ \ \ \ \ \ \ \ \ \ $&$0.9964$\\
$-10\% \ \ \ \ \ \ \ \ \ \ \ \ \ \ \ \ \ \ \ \ \ $&$-10\%\ \ \ \ \ \ \ \ \ \ \ \ \ \ \ \ \ \ \ \ \ $&$ 10\% \ \ \ \ \ \ \ \ \ \ \ \ \ \ \ \ \ \ \ $&$0.9798$\\
$-10\% \ \ \ \ \ \ \ \ \ \ \ \ \ \ \ \ \ \ \ \ \ $&$-10\%\ \ \ \ \ \ \ \ \ \ \ \ \ \ \ \ \ \ \ \ \ $&$ -10\%\ \ \ \ \ \ \ \ \ \ \ \ \ \ \ \ \ \ \ $&$0.9796$\\
\hline \hline
\end{tabular}}
\end{center}

\section{Conclusion}

In conclusion, we have proposed a protocol to prepare W states with
SQUID qubits by using dressed states. Firstly, we examined and
simplified the system's dynamics and obtained the effective
Hamiltonian so that the simplified model can be regarded as a
three-level system. This greatly help us to further investigate
about the speeding up of the system's evolution with dressed states.
Secondly, we applied the method with dressed states to the
simplified three-level model, in order to keep the system evolving
along a suitable dressed state during the evolution. And we
carefully designed the parameters $\theta$, $\dot{\theta}$, $\mu$
and $\dot{\mu}$, which are shown in Eq.~(\ref{e16.1}). With these
parameters, the Rabi frequencies of pulses being designed can be
expressed by the superpositions of Gaussian functions with curve
fitting, so that they are feasible for experimental realization.
Thirdly, we selected a suitable coupling constant $g$ for both
robustness and speediness. With the designed pulses and the chosen
coupling constant, we continued to explore the robustness against
all kinds of influencing factors, including the cavity decay, the
spontaneous emissions of SQUID squbits, the dephasing and some
parameter variations caused by the imperfect operations, and we
found that the present protocol holds great robustness against these
influencing factors. Meanwhile, we compared the evolution speed of
the present protocol with that of STIRAP. The results showed that
the evolution speed of the present protocol is much faster than that
of STIRAP. On the other hand, in experiment, the SQUID qubits have a
lot of advantages as we discussed in Sec. I. Therefore, we hope the
present protocol can be realized in circuit quantum electrodynamics
systems and contribute to the quantum information processing in near
future.

\section*{Acknowledgement}

This work was supported by the National Natural Science Foundation
of China under Grants No. 11575045, No. 11374054 and No. 11675046,
and the Major State Basic Research Development Program of China
under Grant No. 2012CB921601.

\end{document}